\documentclass[12pt]{article}
\usepackage{graphicx,floatflt,amssymb,epsfig,epsf,amsmath} 
\begin{document}
\renewcommand{\thesection}{\Roman{Section}}

\begin{center} 

{\Large \bf Effect of realistic interatomic interactions and two-body correlation on the heat capacity of a trapped BEC.}\\

\vspace*{1cm} 

\renewcommand{\thefootnote}{\fnsymbol{footnote}} 

{\large {\sf Anindya Biswas}\footnote{e-mail : abc.anindya@gmail.com}} \\ 

\vspace{10pt} 

{\small 

   {\em Department of Physics, University of Calcutta, 92 A.P.C. 

        Road, Kolkata 700009, India.}}
\normalsize 

\end{center} 

\begin{abstract}
An approximate hyperspherical many-body theory has been used to calculate the heat capacity and the condensate fraction of a BEC with effective repulsive interaction. The effect of interactions has been analysed and compared with the non-interacting case. It has been found that the repulsive interaction lowers the critical temperature from the value found in the non-interacting case. The difference  between the critical temperatures increases with the increase in the total number of atoms in the trap.
\vskip 5pt \noindent 
\end{abstract} 

\hspace{1cm}
The number density $(n_{d})$ of a typical Bose-Einstein condensate (BEC) is restricted to $10^{12}$ to $10^{14}$ $atoms/cm^{3}$ so that the average interparticle distance is much larger than the range of interatomic interactions. The dilute gas undergoes BEC below a certain critical temperature ($T_c \sim$   nano Kelvin), when most of the atoms go to the single particle ground state. In this state the momenta of the particles are extremely small and the thermal de Broglie wavelength of all particles overlap. The system therefore behaves as a single quantum object. At higher temperatures, atoms get distributed into various low-lying energy levels.\\
\hspace*{1cm}
It is generally stated that a phase transition occurs during the formation of BEC. In contrast with typical classical phase transitions, the origin of this one is the quantum mechanical effects of Bose-Einstein statistics. The discontinuity of the heat capacity or its temperature-derivative is one of the major manifestations in a phase transition. The heat capacity $(C_{A})$ of a system of large but finite number of non-interacting bosons $(A)$ in a three dimensional harmonic trap shows a discontinuity at the transition temperature ($T_{c}^{0}$), in the semi-classical approximation in which the sum over states is replaced by an integral over energy~\cite{dalfovo,pethick}. In this approximation, the chemical potential at temperature $T$, $\mu(T)$, is constant (=its maximum value) for $T\leq T_{c}^{0}$ and decreases suddenly for $T>T_{c}^{0}$. This approximation is valid for a very large number of particles ($A\rightarrow\infty$). Very rapid change of $C_{A}(T)$ near the critical temperature results for the trapped non-interacting bose gas, having a large but finite number  of particles when sums are numerically evaluated~\cite{napolitano}. The rapid change appears to approach a discontinuity as $A\rightarrow\infty$. Thus for a finite system, strictly speaking there is no phase transition, although a rapid change in phase occurs near $T_{c}^{0}$, which is the transition temperature for non-interacting bosons in the thermodynamic limit. Clearly there is no strict transition temperature for the finite system (interacting or non-interacting), a critical temperature ($T_{c}$) may be defined as the temperature at which $C_{A}$ attains its maximum~\cite{napolitano}. In this work we critically examine the nature of these quantities for a finite condensate of trapped interacting bosons.\\
\hspace{1cm}
Interatomic interactions are known to have appreciable effects on the static properties of the condensate~\cite{dalfovo}. Thus it is important to study the effects of two-body interactions on the heat capacity and condensate fraction of the BEC. 
The most common procedure is to solve the Gross-Pitaevskii (GP) equation, which is obtained from the mean field approach, together with the assumption of a contact two-body interaction, whose strength is given by the $s$-wave scattering length ($a_s$)~\cite{dalfovo}. A contact interaction is a good approximation only in the low density limit. Use of a contact interaction in the mean field theory reduces the mean field equation to the $GP$ equation, which is a single second order differential equation, non-linear in the condensate wave function. For a finite range interaction, the ideal procedure would be to solve the many body Schr\"{o}dinger equation \textit{ab initio}. Alternatively, one can use an approximate approach like the fully self-consistent mean field theory. An exact solution of the Schr\"{o}dinger equation is impractical for a large number $(\sim 10^{4})$ of atoms in the condensate. The essentially exact diffusion Monte Carlo (DMC) method~\cite{blume} is a powerful tool for the many-body problem, but it is rather slow and faces difficulties especially for highly excited states of a condensate containing a large number of particles. This is a serious difficulty, since one needs to calculate a large number of excited energy levels of the system to obtain thermodynamic quantities. In this communication, we have adopted an approximate but \textit{ab initio} solution of the many-body Schr\"{o}dinger equation, expanding each Faddeev component of the many-body wave function in a subset [called potential harmonics (PH)]~\cite{fabre} of the full hyperspherical harmonic (HH) basis~\cite{ballot}. The approximation involves disregard of higher-than-two-body correlations in the Faddeev component~\cite{fabre}, which is well justified in a fairly dilute BEC~\cite{das}.\\
\hspace*{1cm}
We have calculated a large number of energy levels ($E_{nl}$) of a condensate of $^{85}Rb$ atoms trapped in a spherically symmetric harmonic oscillator potential and using these the heat capacity and condensate fraction of the system in the condensed as well as the normal phase. Here $E_{nl}$ is the energy in oscillator units of the $n^{th}$ radial excitation of the $l^{th}$ surface mode. The energy eigenstates of the system have been calculated using the Potential Harmonic Expansion Method (PHEM) for trapped interacting bosons~\cite{das,das1}. This technique has been shown to reproduce known results for the static properties~\cite{das,das1,chakrabarti} as also the collapse of attractive condensates~\cite{kundu}. The PHEM was further used to investigate the effect of shape dependence of the two-body potential~\cite{chak}, as also for studying the effect of anharmonic traps~\cite{chak2}. These applications have proved that the underlying methodology of the PHEM produces reasonable results for the $T=0$ properties of the BEC. In the present work, the method is extended to investigate thermodynamic quantities. Convergence of the partition function for $T>0$ requires the calculation of a large number of energy levels (typically $n\sim400$ and $l\sim200$). This is very time consuming for the essentially exact DMC method. Even for the mean field theory and the GP equation, this is a formidable task. By contrast the PHEM is a fairly fast procedure and such a calculation is within the realm of feasibility. \\
\hspace*{1cm}
Here, we consider a system of $A=N+1$ identical bosons, each of mass $m$ and confined in a trap which is approximated by a spherically symmetric harmonic oscillator potential with frequency $\omega$. For the static properties, it is assumed that the atomic cloud is at zero temperature. The time independent Schr\"{o}dinger equation is given by,
\begin{equation}
\left[-\frac{\hbar^{2}}{2m}\Sigma_{i=1}^{A}\nabla_{i}^{2}+ \Sigma_{i=1}^{A}\frac{1}{2}m\omega^{2}x_{i}^{2}+\Sigma_{i,j>i}^{A}V(\vec{x}_{i}-\vec{x}_{j})\right] \Psi(\vec{x})=E'\Psi(\vec{x})
\end{equation} 
where $\vec{x}=\left\lbrace \vec{x}_{1},\vec{x}_{2},......,\vec{x}_{A}\right\rbrace$ represents the position coordinates of $A$ particles, $V(\vec{x}_{i}-\vec{x}_{j})$ is the pairwise local central two-body interaction between the $i^{th}$ and $j^{th}$ particles and $E'$ is the total energy of the system. The centre of mass motion can be decoupled and the Schr\"{o}dinger equation for relative motion of the system is expressed in terms of Jacobi coordinates $\left\lbrace \vec{\xi}_{i} (i=1,N)\right\rbrace$ (which are linear combinations of the position coordinates~\cite{ballot}) as
\begin{equation}
 \left[-\frac{\hbar^{2}}{2m}\Sigma_{i=1}^{N}\nabla_{\xi_{i}}^{2}+\Sigma_{i=1}^{N}\frac{1}{2}m\omega^{2}\xi_{i}^{2}+\Sigma_{i,j>i}^{A}V(\vec{r}_{ij})-E\right] \Psi(\vec{\xi}_{1},\vec{\xi}_{2},...,\vec{\xi}_{N})=0
\end{equation}
where $E=E'-$energy of centre of mass motion and $r=[\Sigma_{i=1}^{N}\xi_{i}^{2}]^{1/2}$ is called the hyperradius. The evolution of the system can be studied by following the motion of one point in the $3N$ dimensional hyperspace. The polar coordinates of this point are given by a set $(\Omega_{N})$ of $(3N-1)$ angles. We choose $\vec{r}_{ij}=\vec{\xi}_{N}$. For the remaining $(N-1)$ Jacobi vectors a hyperradius $\rho_{ij}$ is defined in $3(N-1)$ dimensional space by $\rho_{ij}=[\Sigma_{k=1}^{N-1}\xi_{k}^{2}]^{1/2}.$
In the PHEM, only two-body correlations are incorporated in the wave function. Higher body correlations can be neglected since the gas is very dilute and the probability of a three body collision is minimal. So, the wave function $\Psi(\vec{\xi})$ can be decomposed into Faddeev components
\begin{equation}
 \Psi(\vec{\xi})=\Sigma_{i,j>i}^{A}\psi_{ij}(\vec{r}_{ij},r).
\end{equation}
The Faddeev component $\psi_{ij}$ describes the (partial) motion of the system when the $ij$-pair interacts, while the remaining ($A-2$) particles are simply spectators. The Schr\"{o}dinger equation for the Faddeev component can be written as
\begin{equation}
 \left[-\frac{\hbar^{2}}{2m}\Sigma_{i=1}^{N}\nabla_{\xi_{i}}^{2}+\Sigma_{i=1}^{N}\frac{1}{2}m\omega^{2}\xi_{i}^{2}-E\right] \psi_{ij}(\vec{r}_{ij},r)=-V(r_{ij})\Sigma_{k,l>k}^{A}\psi_{kl}(\vec{r}_{kl},r).
\end{equation}
Summing eq.~$(4)$ over all pairs and using eq.~($3$), we get back the full Schr\"{o}dinger equation. Assumption of two-body correlations alone makes $\psi_{ij}$ a function of $\vec{r}_{ij}$ and $r$ only~\cite{fabre} and hence $\psi_{ij}$ can be expanded in the complete set of potential harmonics (which are the subset of full hyperspherical harmonics, needed for the expansion of $V(r_{ij})$~\cite{fabre}) as
\begin{equation}
 \psi_{ij}(\vec{r}_{ij},r)=r^\frac{-(3A-4)}{2}\Sigma_{K}\textit{P}_{2K+l}^{lm}(\Omega_{ij})u_{K}^{l}(r),
\end{equation}
where $K$ is the grand orbital quantum number and $\Omega_{ij}$ is the set of all hyperangles for the particular choice $\vec{\xi}_{N}=\vec{r}_{ij}$. Substitution of eq.~($5$) in eq.~($4$) and subsequent projection on the PH basis leads to a set of coupled differential equations CDE~\cite{das,das1} which are then solved using the hyperspherical adiabatic approximation (HAA)~\cite{das2,ballot2}. The latter assumes that the hyperradial motion is slow compared to the hyperangular motion. This approximation has been shown to be very reliable in atomic and molecular cases~\cite{das2}. The adiabatically separated hyperangular eigenvalue equation is solved (by diagonalizing the corresponding potential matrix) to obtain the lowest eigenpotential $\omega_{0}(r)$ as a parametric function of $r$. This is the effective potential for the condensate to move as a single quantum entity in the hyperradial space. In the HAA approach, an approximate solution of the CDE is obtained by solving a single uncoupled differential equation,
\begin{equation}
 \left[-\frac{\hbar^{2}}{m}\frac{d^{2}}{dr^{2}}+\omega_{0}(r)+\Sigma_{K=0}^{K_{max}}|\frac{d\chi_{K0}(r)}{dr}|^{2}-E\right]\zeta_{0}(r)=0
\end{equation}
where $\zeta_{0}(r)$ is the condensate wave function in the hyperradial space and $\left\lbrace \chi_{K,0}\right\rbrace$ is the eigen column vector, corresponding to the lowest eigenvalue $\omega_{o}(r)$, of the potential matrix for a fixed value of $r$. Ground state in the effective potential well $\omega_{0}(r)$ gives the ground state energy ($E_{00}$) of the condensate. For the calculation of thermodynamic properties using the grand canonical partition function, we need a large number of excitation levels in this effective potential well which depends on the orbital angular momentum ($l$) of the system. However for $l>0$, computation of the potential matrix element is very time consuming. On the other hand, the hyper-centrifugal repulsion term appearing in the matrix to be diagonalized, is very large for large $A$ compared to the contribution coming from $V(\vec{r}_{ij})$ for $l>0$. Thus the hyper-centrifugal repulsion term contributes most to the full matrix~\cite{biswas}. Hence contributions to the off-diagonal matrix elements arising from $l>0$ are disregarded for the calculation of $E_{nl}$. Contributions coming from all terms for $l=0$ are properly taken~\cite{biswas}. Finally, the hyperradial equation is solved in the extreme adiabatic approximation~\cite{das2} to calculate the ground and excited energy levels of the condensate.\\
\hspace*{1cm}
We perform the calculations for a condensate of $^{85}Rb$ atoms with $a_{s}=2.09\times10^{-4}$ o.u.($6.39\times10^{-10}$ m), which is within the range of values of $a_{s}$ used in the JILA experiment~\cite{cornish}. We select only one typical value of the repulsive $s-$wave scattering length to demonstrate our results, since we need to calculate a large number of energy levels, which is quite time consuming even by the PHEM. More detailed calculations, particularly those for attractive (negative $a_{s}$) condensates will be undertaken later. Although an axially symmetric trap (with radial and axial frequencies $\omega_r$ and $\omega_a$ respectively) was used in the JILA experiment, we assume a spherically symmetric trap of frequency $\omega=(\omega_{r}^{2}\omega_{a})^{\frac{1}{3}}$, for simplicity and to keep our calculations manageable. The interatomic potential is chosen to be a realistic one, \textit{viz.}, the van der Waals potential, with a hard core of radius $r_{c}$
\begin{eqnarray}
V(r_{ij}) & = \infty & ,r_{ij}<r_{c} \nonumber \\
  & = -\frac{C_{6}}{r_{ij}^{6}} & ,r_{ij} \geq r_{c}
\end{eqnarray} 

The value of $C_{6}$ is known for rubidium atoms~\cite{pethick}. Oscillator units (o.u.) are used in our calculations: $\hbar\omega$ for energy and $\sqrt{\frac{\hbar}{m\omega}}$ for length. Value of $C_{6}$ is $6.489755\times10^{-11}$ o.u. ($4.466\times10^{-76} J.m^{6}$). Since the binary collisions occur at extremely low energy, the effective atom-atom interaction is specified by the $s$-wave scattering length $a_{s}$, which in turn depends strongly on $r_{c}$. The zero energy two-body Schr\"{o}dinger equation is solved to obtain $a_{s}$ analytically~\cite{pethick}. The value of $r_{c}$ is adjusted such that $a_{s}$ has the experimental value. Corresponding two-body wave function is used as a short range correlation function for the $PH$ expansion, to enhance its convergence rate~\cite{kundu}. The expansion basis is then truncated subject to the condition of convergence of the $T=0$ static properties of the condensate. Next, a large number of energy levels of the condensate are calculated for each of the orbital angular momenta from $l=0$ to $200$. Calculation of a large number of energy levels is very time consuming. Hence for each value of $l$, a smaller number of low-lying levels were calculated directly solving the hyperradial equation. These were then least square fitted to a suitable power series expansion. Convergence of such an expansion upto the desired accuracy was ascertained. Using this, high-lying levels are then obtained by extrapolation.

\hspace*{1cm}
The Bose distribution function, $f(E_{nl})$, is given by
\begin{equation}
f\left( E_{nl}\right) =\frac{1}{e^{\beta\left( E_{nl}-\mu\right) }-1},
\end{equation}
where $\beta=\dfrac{1}{k_{B}T}$, $k_{B}$ being the Boltzmann's constant, $T$ is the absolute temperature and $\mu\equiv\mu\left( T\right)$ is the chemical potential. Since the number of bosons $(A)$ is fixed, $\mu$ is obtained from the constraint~\cite{napolitano}
\begin{equation}
 A=\Sigma_{l=0}^{\infty}\Sigma_{n=0}^{\infty}d_{l}f(E_{nl}),
\end{equation}
where $d_l=2l+1$ is the degeneracy factor of the $l^{th}$ surface mode. 
The total energy $E(A,T)$ of the system is given by
\begin{equation}
 E(A,T)=\Sigma_{l=0}^{\infty}\Sigma_{n=0}^{\infty}d_{l}f(E_{nl})E_{nl}.
\end{equation}
Sums in eqs.~($9$) and ($10$) are truncated after achieving convergence upto desired accuracy. The heat capacity of the system, $C_{A}(T)$, for fixed particle number ($A$) is given by
\begin{equation}
 C_{A}(T)=\frac{\partial E(A,T)}{\partial T}
\end{equation}
\begin{equation}
=\beta\Sigma_{l=0}^{\infty}\Sigma_{n=0}^{\infty}\frac{d_{l}E_{nl}e^{\beta(E_{nl}-\mu)}}{(e^{\beta(E_{nl}-\mu)}-1)^{2}}\left( \frac{E_{nl}-\mu}{T}+\frac{\partial \mu}{\partial T}\right),
\end{equation}
where $\dfrac{\partial \mu}{\partial T}$ is given by differentiating eq.~$(9)$ with respect to $T$
\begin{equation}
 \frac{\partial \mu}{\partial T}=-\frac{\Sigma_{k=0}^{\infty}\Sigma_{m=0}^{\infty}d_{k}(E_{mk}-\mu)e^{\beta(E_{mk}-\mu)}[f(E_{mk})]^{2}}{T\Sigma_{l=0}^{\infty}\Sigma_{n=0}^{\infty}d_{l}e^{\beta(E_{nl}-\mu)}[f(E_{nl})]^{2}}.
\end{equation}
We look for convergence of the chemical potential, as $(n,l)$ sums are truncated in the double sum in eq.~($9$). This value of $\mu(T)$ is used to calculate $\dfrac{\partial \mu}{\partial T}$ and $C_{A}(T)$, using eqs.~$(13)$ and $(12)$ respectively.\\

\hspace*{.5cm}
\begin{figure}[hbpt]
\vspace{-10pt}
\centerline{
\hspace{-3.3mm}
\rotatebox{0}{\epsfxsize=12cm\epsfbox{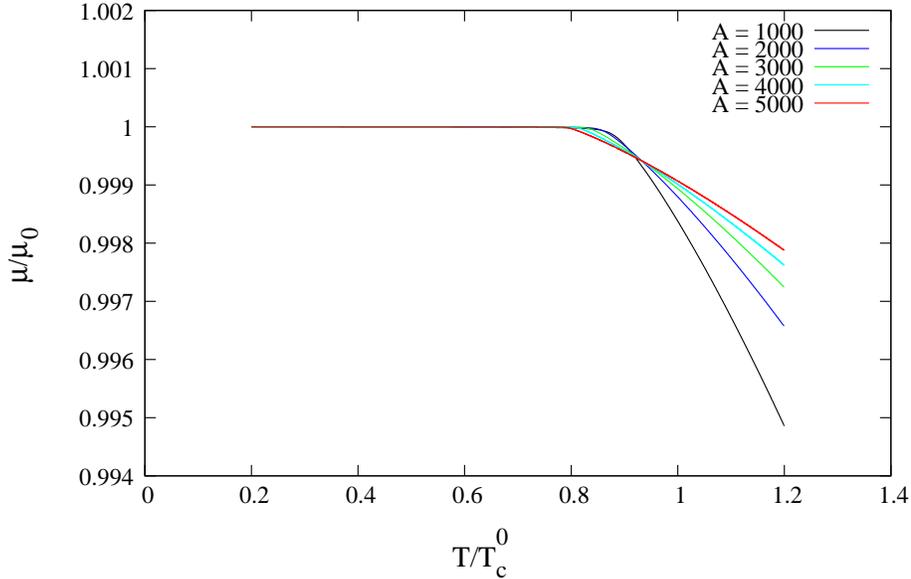}}}
\caption{(Colour online) Chemical potential calculated by PHEM as a function of temperature for indicated number of interacting bosons.}
\end{figure}

\hspace*{1cm}
In Fig.$1$ we plot reduced chemical potential, $\mu / \mu_{0}$ (where $\mu_{0}$ is the chemical potential at $T=0$) as a function of the reduced temperature $T / T_{c}^{0}$ ($T_{c}^{0}$ is the reference critical temperature, $k_{B}T_{c}^{0}=0.94 \hbar \omega A^{1/3}$, according to eq.~($2.20$) of Ref.~\cite{pethick}) for $A = 1000, 2000, 3000, 4000$ and $5000$. Note that in the text book treatment~\cite{pethick}, $\mu$ is taken to be equal to the ground state energy of the system for $T\leq T_{c}^{0}$ and it suddenly starts to differ for $T>T_{c}^{0}$.  In our treatment, since $A$ is relatively small, we evaluate the sums over $n$ and $l$ explicitly and  $\mu(T)$ is determined from the condition ($9$) for all $T>0$. As a consequence $\mu(T)$ is a continuous function of $T$, although $\mu$ remains practically constant over a wide range of $T / T_{c}^{0}$, upto $T / T_{c}^{0}\simeq0.8$. As $T / T_{c}^{0}$ approaches $1$, $\mu / \mu_{0}$ decreases rapidly. Also with increasing $A$ the deviation of $\mu / \mu_{0} $ from $1$ becomes more sudden. It appears that $\dfrac{\partial \mu}{\partial T}$ has a sudden change as $A\rightarrow\infty$.
\hspace*{.5cm}
\begin{figure}[hbpt]
\vspace{-10pt}
\centerline{
\hspace{-3.3mm}
\rotatebox{0}{\epsfxsize=12cm\epsfbox{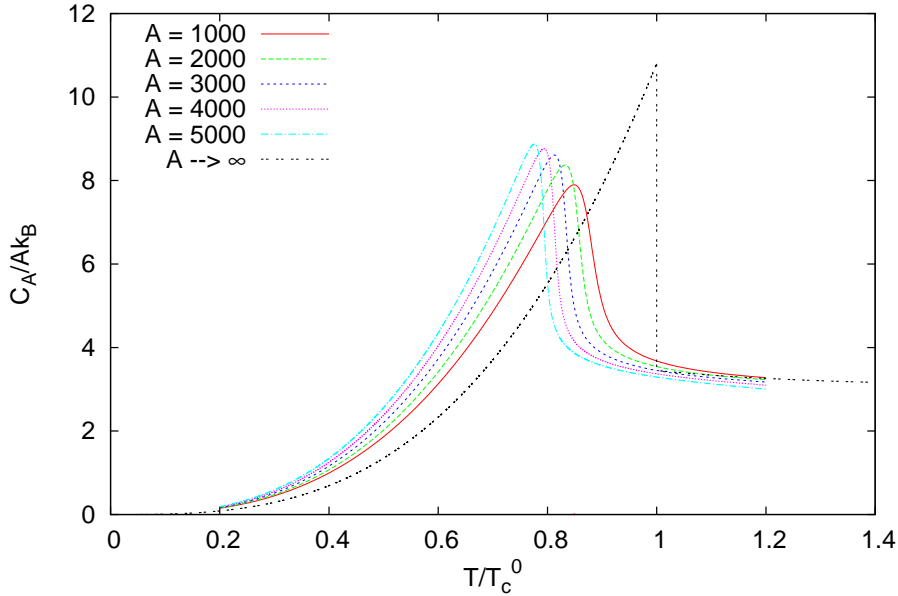}}}
\caption{(Colour online) Heat capacity calculated by PHEM as a function of $T/T_{c}^{0}$ for indicated number of interacting bosons. $A\rightarrow\infty$ indicates infinite number of non-interacting bosons.}
\end{figure}

\hspace*{1cm}
In Fig.$2$, we plot $\dfrac{C_{A}}{Ak_{B}}$ as a function of $T/T_{c}^{0}$ for the same number of particles. The general pattern is similar to the non-interacting case~\cite{napolitano}. One notices a sharp change in $\dfrac{C_{A}}{Ak_{B}}$ just above a critical value ($T_{c}$) of $T$. There is a distinct peak in $C_{A}(T)$. 
In the absence of a discontinuity in $C_{A}$ or its temperature derivative, we follow Ref.~\cite{napolitano} to define the critical temperature ($T_{c}$) to be the temperature at which $\dfrac{\partial C_{A}}{\partial T}=0$. It can be seen from Fig.$2$ that $T/T_{c}^{0}$ for which $\dfrac{C_{A}}{Ak_{B}}$ is maximum decreases with $A$, although $T_{c}$ increases with $A$. Calculated values of $T_{c}$ are comparable with the experimental data~\cite{cornish}. Although a measurement of $T_{c}$ for such a small number of atoms has not been reported, the temperature at which BEC formation was initiated for a larger ($A\gtrsim10000$) number of particles in the trap was reported to be about $15 nK$~\cite{cornish}. The values of $T_{c}$ are listed in table $1$, together with $T_{c}^{0}$ and $T_{c}^{A}$ (critical temperature for a cloud of $A$ non-interacting bosons in an isotropic harmonic trap). It is seen that the effect of interaction lowers the critical temperature. This is similar to the result obtained from the GP equation~\cite{dalfovo}, although the amount of decrease is different. The observation that the critical temperature of the interacting gas decreases compared to the non-interacting atoms is in conformity with other theoretical and experimental findings~\cite{houbiers,gerbier}. The effective repulsive interaction increases the energy of the system; the system therefore has to be cooled to even lower temperatures for all particles to be in the ground state. As $T$ increases above $T_{c}$, most of the atoms get distributed in higher energy levels, with a microscopic fraction of atoms left in the ground state. At $T=T_{c}$, the number of atoms left in the ground level is still appreciable for small $A$ -- it is denoted by $A_{0}(T_{c})$ and presented in the last column in Table $1$. Although $A_{0}(T_{c})$ increases with $A$, the relative fraction $\dfrac{A_{0}(T_{c})}{A}$ decreases with $A$.\\
\hspace*{1cm}
The variation of critical temperature with the number of bosons in the condensate has been presented in Fig.$3$. Dependence of $T_{c}^{0}$, $T_{c}^{A}$ and $T_{c}$ on the number of bosons ($A$) are depicted by curves labelled as $1$, $2$ and $3$ respectively. The interparticle interaction is switched off while calculating $T_{c}^{A}$. The system then effectively reduces to $A$ identical, non-interacting bosons in an isotropic, harmonic potential, which is identical with the calculation of Ref.~\cite{napolitano}. However, both the effects of finite particle number and interparticle interactions are included in the calculation of $T_{c}$. One notices that ($T_{c}^{A}-T_{c}$) increases with $A$. This is intuitively expected since the number of two-body interaction bonds increase as $A(A-1)/2$.\\
\hspace*{.5cm}
\begin{figure}[hbpt]
\vspace{-10pt}
\centerline{
\hspace{-3.3mm}
\rotatebox{-90}{\epsfxsize=8cm\epsfbox{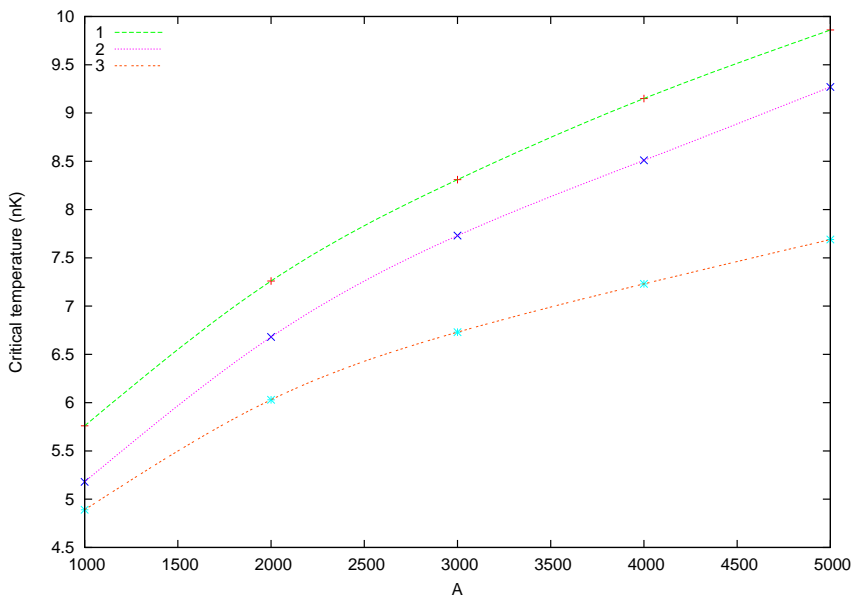}}}
\caption{(Colour online) Critical temperature as a function of number of Bosons confined in a spherically symmetric harmonic oscillator trap : 1 --  non-interacting Bosons in the thermodynamic limit $(T_{c}^{0})$,  2 -- finite number of non-interacting Bosons $(T_{c}^{A})$,  3~--~interacting Bosons $(T_{c})$ by PHEM.}
\end{figure}

\hspace*{.5cm}
\begin{figure}[hbpt]
\vspace{-10pt}
\centerline{
\hspace{-3.3mm}
\rotatebox{0}{\epsfxsize=12cm\epsfbox{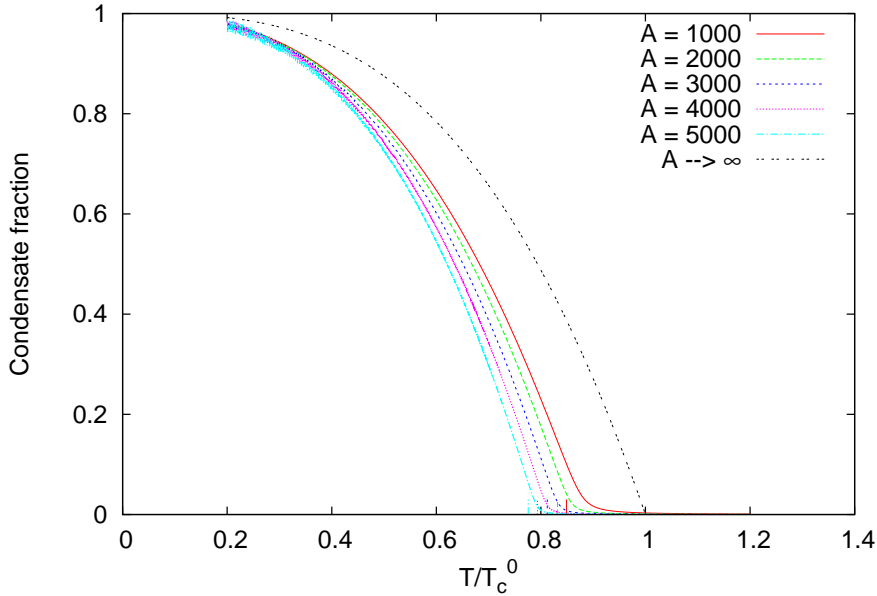}}}
\caption{(Colour online) Condensate fraction calculated by PHEM as a function of $T_{c}/T_{c}^{0}$ for various indicated values of number of interacting bosons. $A\rightarrow\infty$ indicates infinite number of non-interacting bosons.}
\end{figure}

\hspace{1cm}
Finally, we calculate the condensate fraction $(F_{c})$ as 
\begin{equation}
 F_{c}(T)=\frac{A-\Sigma'_{n,l}d_{l}f(E_{nl})}{A}
\end{equation}
where $\Sigma'$ indicates sum over all ($n,l$) except the ground state ($0,0$). We plot it as a function of $T$ in Fig.~$4$ for condensates with $1000$, $2000$, $3000$, $4000$, and $5000$ particles. These plots are again similar to those for the ideal non-interacting case~\cite{dalfovo,pethick}. Some fluctuations, especially for larger number of  particles at lower temperatures, are due to numerical errors. In the same figure, we also plot the condensate fraction of non-interacting bosons in the thermodynamic limit (indicated by $A\rightarrow\infty$). While $F_{c}(T)$ for the non-interacting system in the thermodynamic limit reaches zero sharply at $T=T_{c}^{0}$, that for the finite interacting system decreases fairly gradually for $T>T_{c}$, after a sharp drop at $\simeq T_{c}$. Also as net interaction increases due to increase in $A$, the curves are shifted further to the left, in agreement with the shift of $C_{A}(T)$ (Fig.$2$). In Fig.$5$, we compare the condensate fraction by the PHEM with that obtained using other approaches~\cite{dalfovo,xiong}. The curves are plotted for a particular value of the dimensionless interaction parameter $\eta=\dfrac{\mu(N,T=0)}{k_{B}T_{c}^{0}}$. The parameter $\eta$ is a measure of the ratio of interaction energy and the thermal energy~\cite{dalfovo}. The larger the value of $\eta$, the larger is the interaction energy. The PHEM result for $A=5000$ corresponds to $\eta=0.24$ and is plotted against the reduced temperature $T/T_{c}^{0}$, labelled 'PHEM'. For comparison with the mean field local density result, we use eq.~($122$) of Ref.~\cite{dalfovo} for $\eta=0.24$ to plot the curve labelled 'GP'. The effect of two-body interaction reduces the condensate fraction appreciably. The condensate fraction in a BEC with interacting bosons has also been calculated using the canonical ensemble~\cite{xiong} (eq.~($45$) of Ref.~\cite{xiong}) and labelled 'canonical' in Fig.~$5$. In the same figure we also plot $F_{c}(T)$ for non-interacting bosons in the thermodynamic limit which is labelled 'non-int'. The thermodynamical properties calculated using the canonical ensemble coincides with those obtained using the grand canonical ensemble in the thermodynamic limit. However, the effect of finite number of particles has been incorporated in the curve marked 'canonical' and therefore differs from the curve obtained using the mean field, local density approach in the grand canonical ensemble. Contact interaction has been used for both these approaches. In our method we use the realistic van der Waals potential, incorporate two-body correlations in the condensate wave-function and compute thermodynamical properties for finite number of interacting bosons. The difference of our results from those obtained using the other approaches can be attributed to these causes.\\

\hspace*{.5cm}
\begin{figure}[hbpt]
\vspace{-10pt}
\centerline{
\hspace{-3.3mm}
\rotatebox{0}{\epsfxsize=12cm\epsfbox{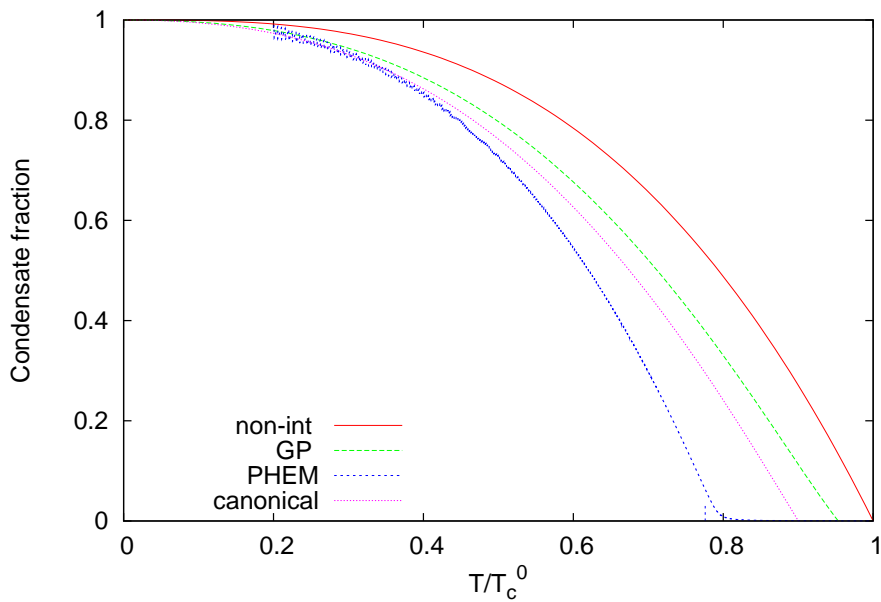}}}
\caption{(Colour online) Calculated condensate fraction as a function of $T/T_{c}^{0}$ for the dimensionless parameter $\eta=0.24$, by different indicated methods.}
\end{figure}

\begin{table}
\caption{Critical temperatures in $nK$ for different values of $A$ and the remaining number of bosons in the lowest energy state at $T=T_{c}$.}
\begin{center}
\begin{tabular}{l|l|l|l|l}
$A$ & $T_{c}$ & $T_{c}^{0}$ & $T_{c}^{A}$ & $A_{0}(T_{c})$\\
\hline
1000 & 4.89 & 5.76 & 5.18 & 102 \\ 
2000 & 6.04 & 7.26 & 6.68 & 168 \\ 
3000 & 6.75 & 8.31 & 7.73 & 225 \\ 
4000 & 7.26 & 9.15 & 8.51 & 268 \\ 
5000 & 7.65 & 9.86 & 9.27 & 317 \\
\end{tabular}
\end{center}
\end{table}
\hspace{1cm}
To summarise, we have calculated the chemical potential, condensate fraction and heat capacity of a condensate containing a fixed number of atoms as a function of temperature (T), using static energy levels calculated by the potential harmonic expansion method. A realistic interatomic interaction $viz.$, van der Waals potential (whose short range behaviour is adjusted to give the correct experimental $s$-wave scattering length) is used as the two-body interaction. In this hyperspherical many-body method all the two-body correlations are appropriately taken care of, but higher-than-two-body correlations are disregarded, which is justified for the dilute condensate. Calculated $C_{A}$ shows a gradual increase with T, until it reaches a maximum and falls rapidly near the critical temperature ($T_{c}$). There is no discontinuity either in $C_{A}$ or its temperature derivative as functions of $T$. The sharpness of the sudden fall increases with $A$. This is similar to the non-interacting inhomogeneous BEC where $C_{A}$ appears to have a discontinuity at $T=T_{c}^{0}$ as $A\rightarrow\infty$ ($T_{c}^{0}$ is the critical temperature in the thermodynamic limit). Beyond $T_{c}$, $C_{A}/A$ approaches the ideal value $3k_{B}$. We notice that  $T_{c}$ increases gradually with $A$, which is also seen for non-interacting atoms. We find that critical temperature for interacing atoms is lower than that of non-interacting atoms, which agrees with intuitive expectations.\\
\hspace{1cm}
Financial support from University Grants Commission (UGC), India is gratefully acknowledged. The author wishes to thank Prof. T. K. Das and Dr. B. Chakrabarti for useful discussions.\\


\newpage
\begin{thebibliography}{References:}
\bibitem{dalfovo} F. Dalfovo et al, {Rev. Mod. Phys. {\bf{71}},  463 (1999)}.
\bibitem{pethick} C. J. Pethick and H. Smith, {{\it Bose-Einstein condensation in dilute gases}, Cambridge, England, 2002}.
\bibitem{napolitano} R. Napolitano, J. De Luca and V.S.Bagnato, {Phys. Rev. A {\bf 55},  3954 (1997)}.
\bibitem{blume} D. Blume and C. H. Greene, {Phys. Rev. A {\bf 63},  063601 (2001)}.
\bibitem{fabre} M.Fabre de la Ripelle, {Ann. Phys. (N.Y.) {\bf 147}, 281 (1983)}.
\bibitem{ballot} J.L.Ballot and M.Fabre de la Ripelle, {Ann. Phys. (N.Y.) {\bf 127}, 62 (1980)}.
\bibitem{das} T.K.Das and B.Chakrabarti, {Phys. Rev. A {\bf 70}, 063601 (2004)}.
\bibitem{das1} T.K.Das et al, {Phys. Rev. A {\bf 75}, 042705 (2007)}.
\bibitem{chakrabarti} B.Chakrabarti, A.Kundu and T.K.Das, {J. Phys. B {\bf 38}, 2457 (2005)}.
\bibitem{kundu} A.Kundu, B.Chakrabarti and T.K.Das, {J. Phys. B {\bf 40}, 2225 (2007)}.
\bibitem{chak} B. Chakrabarti and T. K. Das, {Phys. Rev. A {\bf 78}, 063608 (2008)}.
\bibitem{chak2} B. Chakrabarti, T. K. Das, and P. K. Debnath, {Phys. Rev. A {\bf 79}, 053629 (2009)}.
\bibitem{das2} T.K.Das, H.T.Coelho and M.Fabre de la Ripelle, {Phys. Rev. C {\bf 26}, 2281 (1982)}.
\bibitem{ballot2} J.L.Ballot, M. Fabre de la Ripelle and J.S.Levinger, {Phys. Rev. C, {\bf{26}}, 2301 (1982)}.
\bibitem{biswas} A. Biswas and T.K.Das, {J. Phys. B {\bf 41}, 231001 (2008)}.
\bibitem{cornish} S.L.Cornish et al, {Phys. Rev. Lett. {\bf 85}, 1795 (2000)}.
\bibitem{houbiers} M. Houbiers, H. T. C. Stoof, and E. A. Cornell {Phys. Rev. A {\bf 56}, 2041 (1997)}.
\bibitem{gerbier} F. Gerbier et al, {Phys. Rev. Lett. {\bf 92}, 030405 (2004)}.
\bibitem{xiong} H. Xiong et al, {Phys. Rev. A {\bf 65}, 033609 (2002)}.
\end{thebibliography}
\end{document}